\documentstyle[graphicx]{mn}

\title{Do LINER 2 galaxies harbour low-luminosity Active Galactic Nuclei?}

\author[T.\,P. Roberts, N.\,J. Schurch and R.\,S. Warwick ]
{T.\,P. Roberts, N.\,J. Schurch and R.\,S. Warwick \\
Department of Physics and Astronomy, University 
of Leicester, University Road, Leicester, LE1 7RH}
\date{}

\def\ro{{\it ROSAT~\/}}
\def\asca{{\it ASCA~\/}}
\def\ein{{\it EINSTEIN~\/}}
\def\xmm{{\it XMM-NEWTON~\/}}

\def\hst{{\it HST~\/}}
\def\chan{{\it CHANDRA~\/}}

\def\Msun{\hbox{$\rm ~M_{\odot}$}}

\def\ergsec{{\rm ~erg~s^{-1}}}

\def\atpcm{{\rm ~atoms~cm^{-2}}}
\def\ctsec{{\rm ~count~s^{-1}}}

\def\H0{{\rm ~km~s^{-1}~Mpc^{-1}}}

\def\eg{{\it e.g.~\/}}

\def\ie{{\it i.e.~\/}}
\def\cf{{\it c.f.~\/}}
\def\la{\mathrel{\hbox{\rlap{\hbox{\lower4pt\hbox{$\sim$}}}{\raise2pt\hbox{$<$}}}}}
\def\ga{\mathrel{\hbox{\rlap{\hbox{\lower4pt\hbox{$\sim$}}}{\raise2pt\hbox{$>$}}}}}

\def\d25{D$_{25}$}
\def\nh{{$N_{H}$}}
\def\Ha{{H$\alpha$}}
\def\hi{H {\small I}$~$}
\def\hii{H {\small II}$~$}

\def\.25{0.25 keV\thinspace}
\def\lx{L$_{\rm X}$}

\begin{document}

\maketitle

\begin{abstract}
We use \ro HRI spatial data and \asca spectral measurements for a
sample of seven nearby, early type spiral galaxies, to address the
question of whether a low-luminosity Active Galactic Nucleus (LLAGN)
is present in galaxies that have a LINER 2 classification.  The
brightest discrete X-ray source in the \ro HRI observations is
invariably found to be positionally coincident with the optical
galactic nucleus, and in most cases its flux dominates the X-ray
emission from the central region of the galaxy.  All seven galaxies
have X-ray spectra consistent with a two-component, soft thermal plus
hard power-law, spectral form. If we exclude the two galaxies with
relatively hard X-ray spectra, NGC 3628 and NGC 4594, for which there
is supporting evidence for a LLAGN (or alternatively in the case of
NGC 3628 a dominant ultra-luminous X-ray binary), then the remaining
galaxies show surprisingly similar X-ray spectral
properties. Specifically the flux ratio $F_X$(0.5-1)/$F_X$(2-5), which
measures the relative strengths of the thermal and non-thermal
emission components, shows little scatter about a mean of 0.66, a
value very similar to that measured in the classic starburst galaxy
NGC 253. Since there is no obvious reason why the luminosity of the
hard power-law continuum emanating from a putative LLAGN should be
very closely correlated with the thermal emission of the surrounding
region, this suggests that that the broad-band (0.5-5 keV) X-ray
emission from these LINER 2 galaxies may originate in a common set of
processes probably associated with the starburst
phenomenon. Conversely, it appears that in many, perhaps the majority,
of LINER 2 galaxies, the nuclear X-ray luminosity does not derive
directly from the presence of a LLAGN.

\end{abstract}


\section{Introduction}

The widespread nature of the low-ionization nuclear emission-line
region (LINER) phenomenon (Heckman 1980) has recently been underlined
by Ho, Filippenko \& Sargent (1997, hereafter HFS), who find that
galaxies containing LINERs comprise $\sim 33$\% of the local bright
galaxy population, outnumbering low-luminosity Seyfert nuclei by a
ratio of 3:1.  The excitation mechanisms giving rise to the
characteristic optical spectra of LINERS remain the subject of
considerable debate.  Plausible models include photoionization by a
non-stellar continuum source, namely a low-luminosity active galactic
nucleus (LLAGN), collisional ionization by fast shocks (Heckman 1980)
and photo-ionization by the UV radiation generated by clusters of hot,
young stars (Terlevich \& Melnick 1985, Shields 1992).

The detection by HFS of broad ($\sim$ few thousand km s$^{-1}$)
permitted lines in the optical spectra of $\sim 20\%$ of LINERs points
firmly to the presence of a LLAGN in a subset of the population.  This
also prompted HFS to introduce a ``type 1'' and ``type 2''
classification for LINERS, paralleling the established nomenclature
for Seyfert galaxies.  In light of the fact that many nearby galaxies
appear to contain a central supermassive black hole (see Lawrence 1999
for a review), it is a reasonable conjecture that the majority of
LINERs (i.e., all type 1 plus a significant fraction of type 2
systems) may be powered by a LLAGN. Such a conclusion has consequences
for a range of astrophysical issues. It would, for example, imply that
LINERs constitute the low-luminosity end of the local AGN population
with ramifications for the evolution of the AGN luminosity function
and the history of supermassive black hole formation in galaxies.

X-ray observations can provide important clues as to the physical
processes responsible for LINERs.  Ideally, with good spatial
resolution, extended nebulosity can be resolved to reveal the presence
(or absence) of a compact X-ray source identifiable as the LINER
nucleus. With moderate spectral resolution a non-thermal AGN-like
spectrum is then readily distinguished from a thermal (hot gas)
spectrum. Even when the spatial resolution fails to separate all the
components, spectral modelling may allow the various contributors to
the composite spectrum to be distinguished.  Furthermore, since X-ray
observations can be used to detect a LLAGN which is inconspicuous in
the optical band (either due to its low contrast against the stellar
backdrop or the effects of nuclear obscuration), X-ray measurements
provide an effective way of testing the hypothesis that the majority
of LINER 2's are accretion-powered systems.

In the present paper we focus our attention on a small sample of
nearby galaxies classified by HFS as having either a pure LINER 2
optical spectrum or a composite LINER 2/HII spectrum. HFS refer to the
latter as transition LINERs. For all the sources in our sample, X-ray
spectral information is available from \asca and high spatial
resolution X-ray imaging from \ro.

\section{The LINER 2 sample}

The LINER 2 galaxies studied in this paper were selected primarily on
the basis of the availability of appropiate \asca and \ro High
Resolution Imager (HRI) datasets. As a first step, the positions of
the 133 galaxies classified as LINER 2 (L2) or Transition 2 (T2)
objects by HFS were cross-correlated with the pointing positions of
the ASCA observations comprising the ASCAPUBLIC database available at
the Leicester Database and Archive Service (LEDAS)\footnote{We also
include a proprietary observation of NGC 3627 in our sample.}.  This
lead to a preliminary list of 31 L2/T2 galaxies.  This sample was
further reduced by eliminating all the galaxies that were not the
target of the \asca observations. Also at this stage elliptical and
lenticular galaxies were excluded from the sample, due to concern
that X-ray emission associated with a hot ISM component in these
galaxies might confuse the search for nuclear X-ray activity. Discarding 
the only galaxy of the 8 remaining galaxies that has not been observed 
with the \ro HRI instrument, results in a small homogeneous sample of 7 
early-type spiral galaxies. Basic details of these seven galaxies, 
taken from HFS, are given in Table~\ref{galaxies}. Also the specific \asca
and \ro HRI datasets used in the present analysis are listed in
Table~\ref{alldata}. Information on the LINER nucleus and the X-ray properties
of each galaxy is summarised in the Appendix.

\begin{table*}
\centering
\caption{Properties of the galaxies in the sample.}
\begin{tabular}{lccccc}
Name	& Nuclear class	& $T$	& Hubble type	& Distance (Mpc)
& M$_{\rm B, bulge}$ \\ 
 & & & & & \\
NGC 3627	& T2/S2	& 3	& SAB(s)b		& 6.6	& -18.43 \\
NGC 3628	& T2		& 3	& SAb pec spin	& 7.7	& -18.58 \\
NGC 4321	& T2		& 4	& SAB(s)bc	& 16.8	& -19.18 \\
NGC 4569	& T2		& 2	& SAB(rs)ab	& 16.8	& -20.11 \\
NGC 4594	& L2		& 1	& SA(s)a spin	& 20.0	& -22.11 \\
NGC 4736	& L2		& 2	& (R)SA(r)ab	& 4.3	& -18.19 \\
NGC 7217	& L2		& 2	& (R)SA(r)ab	& 16.0	& -19.26 \\
\end{tabular}
\label{galaxies}
\end{table*}

\begin{table*}
\centering
\caption{The \ro and \asca datasets used in the present analysis.}
\begin{tabular}{lccccccccc}
Galaxy & & & \multicolumn{2}{c}{\ro data} & & & \multicolumn{3}{c}{\asca data}
 \\ 
 & & & ID & Exposure (s) & & & ID & ~~~SIS exp.(s) & GIS exp.(s) \\ 
 & & & & & & & & & \\
NGC 3627	& & & rh600836a01	& 9456 		& & & 67031000
& 19345 & 24688 \\ 
NGC 3628	& & & rh700009n00	& 13574 	& & & 61015000
& 23904 & 26912 \\ 
NGC 4321	& & & rh600731n00	& 42797 	& & & 55044000
& 27296 & 30336 \\ 
NGC 4569	& & & rh600603a01	& 21858 	& & & 65012000
& 19520 & 21728 \\ 
		& & &			&		& & & 65012010
& 19424 & 20960 \\ 
NGC 4594	& & & rh600044a02	& 15850 	& & & 61014000
& 19552 & 20864 \\ 
NGC 4736	& & & rh600678n00	& 112910 	& & & 63020000
& 28032 & 32736 \\ 
NGC 7217	& & & rh702933a01	& 10791 	& & & 63009000
& 79552 & 86624 \\ 
\end{tabular}
\label{alldata}
\end{table*}

\section{The \ro HRI observations}

The \ro HRI observations have been used to examine the X-ray morphology
of each of the galaxies in our sample.  The HRI measurements cover a
soft (0.1 -- 2.4 keV) energy range and lack the sensitivity to
highly obscured (\nh $> 10^{22.5} \rm~cm^{-2}$) sources afforded by
the harder (0.6 -- 8 keV) effective bandpass of \asca. However, the
former has a massive advantage over the latter in terms of spatial
resolution with a point spread function FWHM of $\sim 5''$ compared to
$\sim 90''$.  Although some previous galaxy studies (e.g.  Ptak et
al. 1997; Terashima et al. 2000a) have attempted to utilise the imaging
capabilities of \asca, the relatively low count rates combined with
the inherent poor spatial resolution dissuades us from adopting this
approach in the present study.  Here we focus solely on the spatial
information available from the \ro HRI observations.

We have analysed the deepest \ro HRI observation available for each
galaxy (see Table~\ref{alldata}).  The spatial analysis was performed
in two stages.  Firstly, the raw HRI images were smoothed using a
Gaussian mask with FWHM of 15$''$, and the corresponding X-ray contour
maps were overlaid onto Digitised Palomar Sky Survey images as shown
in Figure 1.  Secondly, a discrete source search was
undertaken using the {\small PSS} (Point Source Search) algorithm, as
described in Roberts \& Warwick (2000). All the X-ray sources having a
statistical significance of $5\sigma$ or greater lying within the \d25
isophotal ellipse of each galaxy were recorded.

\vspace{5mm}

\textbf{Insert Figure 1 here.}

\vspace{5mm}

\setcounter{figure}{1}


The most striking feature of the images in Figure 1 is that
the X-ray emission coincident with the each galactic disk appears to
peak at the position of the nucleus of the galaxy. This result is
verified by the {\small PSS}, which detects a point-like source within
$\sim 30''$ of the optical nuclear position in every image. These
apparently ubiquitous nuclear X-ray sources are, without exception,
the most luminous point-like X-ray source associated with each
galaxy. The observed HRI count rate of the point-like nuclear
component, $C_{nuc}$, and the corresponding (Galactic
absorption-corrected) X-ray luminosity, \lx, is given in
Table~\ref{hrinxs}.  Here we derive X-ray luminosity (in the 0.1--2.4
keV band) using the same spectral assumptions as employed by Roberts
\& Warwick (2000).

Table~\ref{hrinxs} also lists the fraction, $f_{nuc}$, of the counts
measured in the HRI image within a radius of 4$'$ of the nucleus,
which are attributable to the point-like nuclear X-ray source.  (A
radius of 4$'$ was chosen to match the source extraction region used
in the \asca analysis.)  Clearly $f_{nuc}$ is an indicator of how
dominant the nuclear X-ray source is in comparison to the other
contributors (both extended and point-like) to the central X-ray
emission from each galaxy. For the majority of the sources the nuclear
point component represents a substantial fraction of the total central
X-ray emission, with NGC 3627 at $\sim 15\%$ having the lowest
fractional nuclear contribution.

\begin{table*}
\centering
\caption{Nuclear X-ray sources detected in the HRI images.}
\begin{minipage}{150mm}
\begin{tabular}{lccc}
Source	& $C_{nuc}~^{a}$ & log $L_X~^{b}$	&  $f_{nuc}$ \\
& & & \\
NGC 3627  & 3.9 $\pm$ 0.8  & 39.00 & $0.15^{+0.05}_{-0.04}$ \\
NGC 3628  & 12.3 $\pm$ 1.0 & 39.62 & $1.00^{+0}_{-0.08}$  \\ 
NGC 4321  & 8.2 $\pm$ 0.5  & 40.13 & $0.42^{+0.06}_{-0.05}$  \\ 
NGC 4569  & 5.1 $\pm$ 0.6  & 39.93 & $0.51^{+0.14}_{-0.11}$  \\ 
NGC 4594  & 17.3 $\pm$ 1.2 & 40.65 & $0.51 \pm 0.07$  \\ 
NGC 4736  & 50.5 $\pm$ 0.7 & 39.69 & $0.62 \pm 0.02$  \\ 
NGC 7217  & 2.3 $\pm$ 0.6  & 39.71 & $0.39^{+0.25}_{-0.16}$  \\ 
\end{tabular}
\footnotetext{$^{a} \rm In~units~of~10^{-3}~HRI~count~s^{-1} $}
\footnotetext{$^{b} \rm In~erg~s^{-1}~~(0.1-2.4~keV)$}
\end{minipage}
\label{hrinxs}
\end{table*}

To the best of our knowledge the \ro HRI observations of NGC 3627 and
NGC 7217 are previously unpublished. NGC 3627 is particularly
interesting since its nuclear emission is relatively weak (\cf Table
3) with the bulk of the central soft X-ray emission originating in an
irregular diffuse component with an angular scale of $\sim 3'$.  The
northern tip of this extended component overlaps with the position of
the recent supernovae SN1989B.  Two extra-nuclear point-sources are
detected (to the east of nucleus) with X-ray luminosities (corrected
for Galactic absorption) of $4.5$ and $9 \times 10^{38} \ergsec$
respectively. These may be two further examples of ``super-luminous
objects'', a heterogeneous class of very luminous non-nuclear X-ray
sources which have been observed in a number of nearby galaxies
(Roberts \& Warwick 2000). NGC 3627 is listed as an AGN candidate by
Filho, Barthel \& Ho (2000) on the basis of their VLA
observations. This together with the ambiguous T2/S2 spectral
classification assigned by HFS might be explained if NGC 3627 hosts a
Seyfert 2-like nucleus, which is heavy absorbed along the line of
sight. In this scenario the observed X-ray luminosity could arise as
nuclear continuum flux which is electron scattered into the line of
sight by highly photoionized gas extending above (and below) the
obscuring matter.

\section{The ASCA observations}

The details of the ASCA SIS and GIS datasets used in the present
analysis are given in Table~\ref{alldata}. Bright-2 mode data were
available for all the sources with the exception of NGC 3628 for which
only Bright mode data were available. The data were analysed using
standard procedures and data-screening criteria. Images from each
\asca instrument were examined and raw spectra and light curves
extracted, using a circular aperture of radius 4$'$ centred on the
source position.  Complementary off-source (i.e. background)
information was also obtained for each dataset, and subtracted in the 
spectral fitting process. The net source count rates were in the range 
$0.009 - 0.85 \ctsec$ for the SIS and $0.006 - 0.55 \ctsec$ for the GIS.  
In each case the analysis concentrated on the temporal and spectral 
properties of the target galaxy.

In the case of the temporal analysis, the data from each \asca
telescope were first background subtracted and then binned on a
timescale of a quarter of an \asca orbit (1440 seconds). The resulting
light curves showed no evidence for significant variations on any
sampled timescale. In the case of NGC 4569 two separate observations
were available but again no significant variations were apparent
between the two datasets.

The response matrices and ancillary response files for the spectral
investigation of the SIS data were created using the standard \asca
{\small FTOOLS}.  The most recent standard response matrices and
ancillary response files were used for the GIS instrument
analysis. The observation-averaged and background-subtracted spectral
data were binned to a minimum of 20 counts per spectral channel over
the range 0.6 - 10 keV for the SIS and 0.8 - 10 keV for the GIS
instruments.

\begin{table*}
\centering
\caption{Results of the spectral fitting of the \asca observations.}
\begin{minipage}{150mm}
\begin{tabular}{lccccccc}
Source	& $kT$ (keV)	& N$_{H,Gal}~^{a}$	& N$_{H,Int}~^{b}$	& $\Gamma$	&
$\chi$$^{2}$/d.o.f.	& $\Delta \chi^{2}$     & log $L_X~^{c}$\\
& & & & & & \\
NGC 3627	& 0.88$^{+0.02}_{-0.07}$	& 2.4 & 1.7$^{+0.82}_{-0.63}$
& 2.6$^{+0.38}_{-0.37}$      & 196.6/202	& 16.9	& 39.81 \\ 	
NGC 3628	& 0.85$^{+\infty}_{-0.30}$	& 2.2 & 1.5$^{+0.79}_{-0.70}$
& 2.0$^{+0.40}_{-0.38}$      & 89.6/85	& 11.0	& 39.85 \\ 
NGC 4321	& 0.67$^{+0.66}_{-0.67}$	& 2.4 & 0.2$^{+1.54}_{-0.20}$
& 1.9$^{+0.56}_{-0.48}$      & 77.9/91	& 1.6	& 40.33\\ 
NGC 4569	& 0.65$^{+0.09}_{-0.10}$	& 2.5 & 2.3$^{+1.27}_{-0.73}$
& 2.6$^{+0.61}_{-0.53}$      & 155.2/133	& 48.5	& 40.34\\ 
NGC 4594	& 0.76$^{+0.07}_{-0.17}$	& 3.8 & 0.8$^{+0.44}_{-0.37}$
& 1.96$^{+0.14}_{-0.16}$   & 272/239	& 30.6	& 41.19 \\ 
NGC 4736	& 0.64$^{+0.05}_{-0.07}$	& 1.44 & 0.03$^{+0.12}_{-0.30}$
& 1.60$^{+0.15}_{-0.08}$   & 277.6/281	& 201.3	& 39.53 \\ 
NGC 7217	& 0.70$^{+0.13}_{-0.23}$ 	& 0.11 & 0.64$^{+0.70}_{-0.64}$
& 1.8$^{+0.44}_{-0.38}$    & 146.7/168	& 13.0  & 40.10 \\  
\end{tabular}
\footnotetext{$^{a} \rm In~units~of~10^{20}~cm^{-2}. Fixed.  $}
\footnotetext{$^{b} \rm In~units~of~10^{22}~cm^{-2}   $}
\footnotetext{$^{c} \rm In~erg~s^{-1}~(0.5-10~keV)$}
\end{minipage}
\label{bestfits}
\end{table*}

\begin{table*}
\centering
\caption{90\% upper limits to the line equivalent width for various 
iron K$\alpha$ configurations.}
\begin{minipage}{150mm}
\begin{tabular}{lcccc}
Source & 6.4 keV line    & 6.7 keV line  & Free line & Free line \\
       & Eq.Width (eV)   & Eq.Width (eV) & Energy (keV) & Eq.Width (eV)  \\
& & & & \\
NGC 3627 & 467 & 710 & - & - \\
NGC 3628 & 694 & 859 & - & - \\
NGC 4321 & 670 & 804 & - & - \\
NGC 4569 & 655 & 486 & - & - \\
NGC 4594 & 205 & 316 & - & - \\
NGC 4736 & 350 & 604 & 6.72$^a$ & 597 \\
NGC 7217 & 606 & 1320 & $6.94^{+0.26}_{-0.64}$ & 1688 \\
\end{tabular}
\footnotetext{$^a$ Line energy unconstrained at the 90\% level}
\end{minipage}
\label{iron}
\end{table*}

\begin{figure*}
\centering
\hbox{
\includegraphics[width=5.5 cm,angle=270]{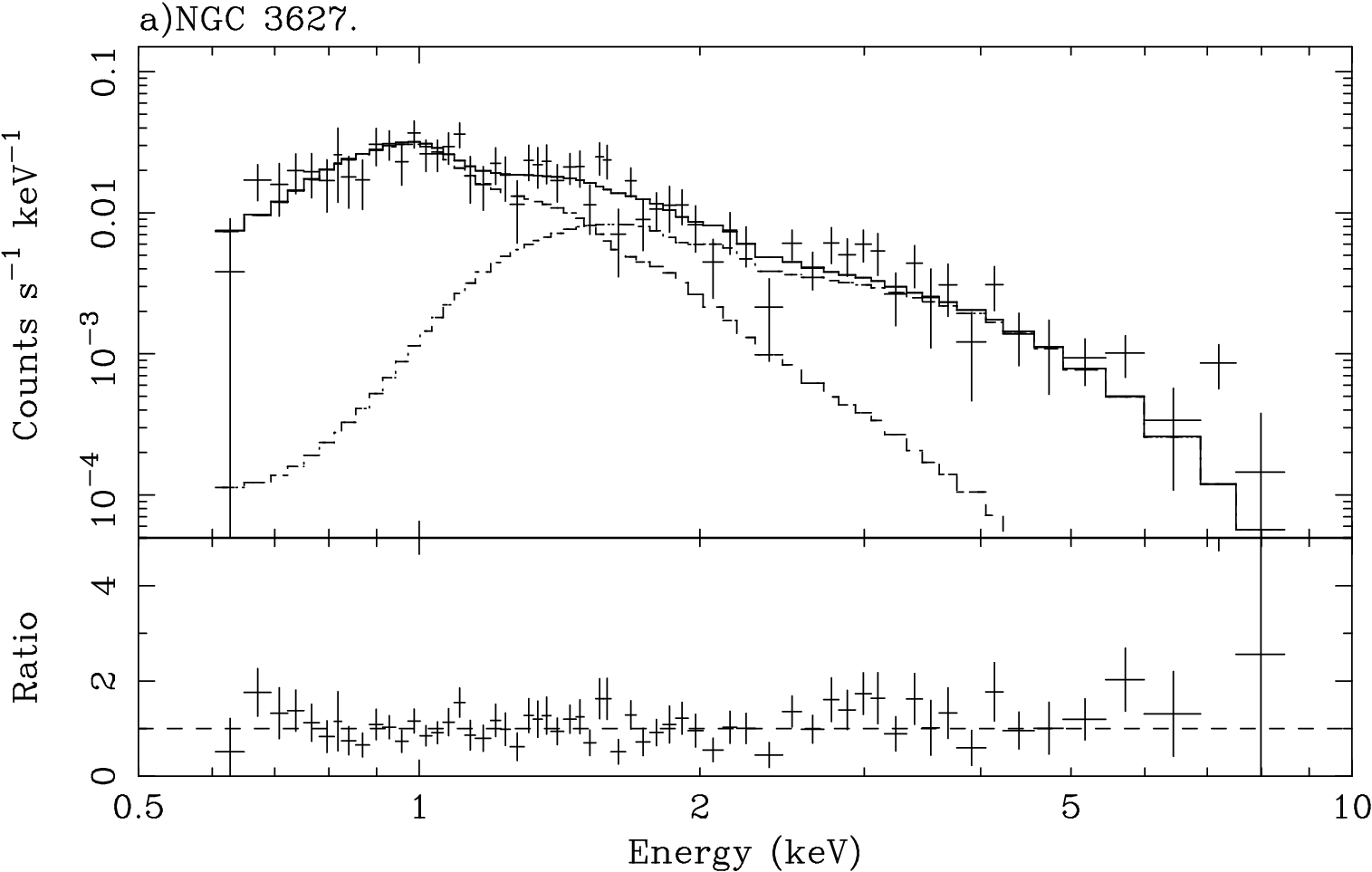}
\includegraphics[width=5.5 cm,angle=270]{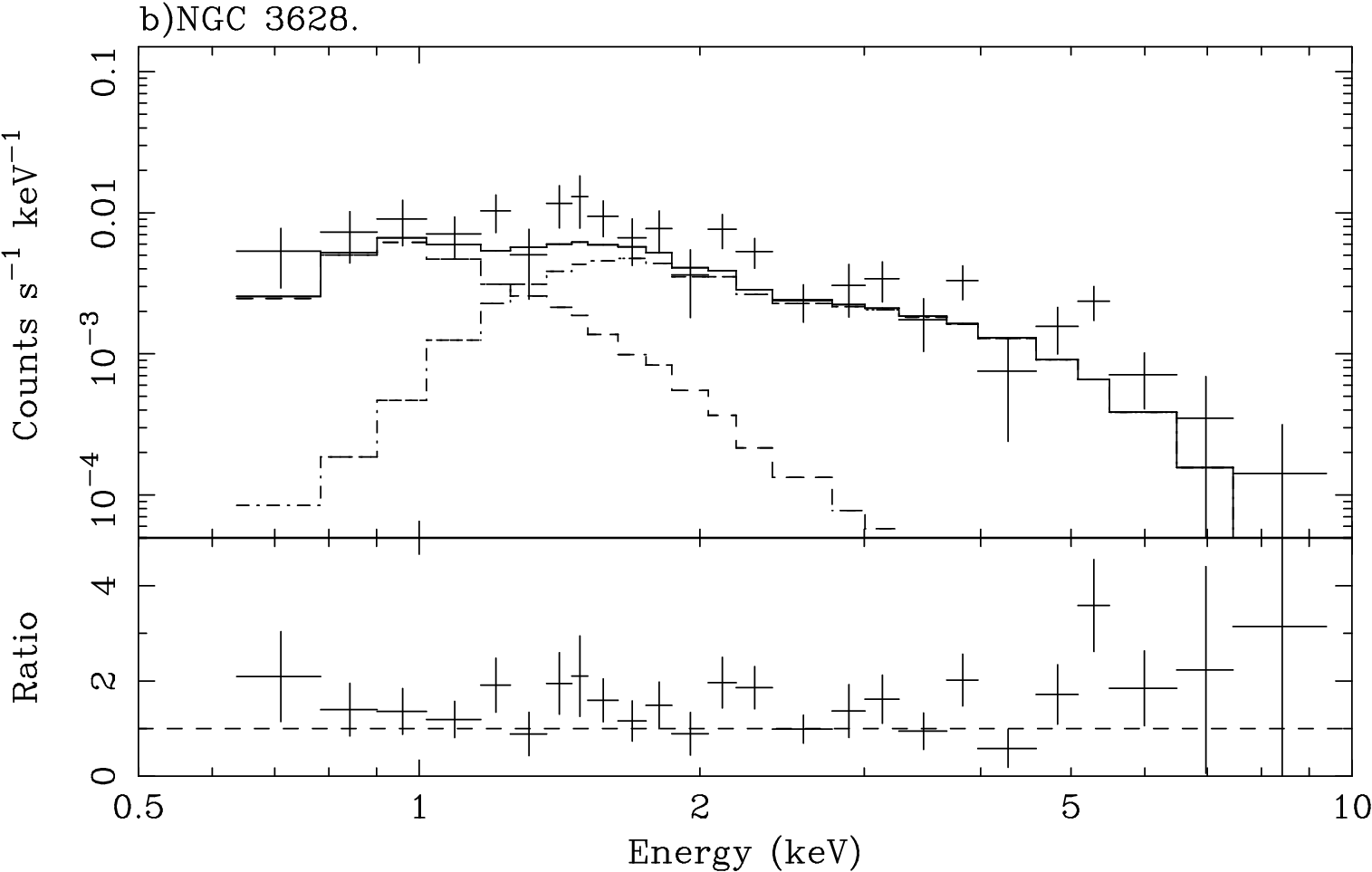}
}
\hbox{
\includegraphics[width=5.5 cm,angle=270]{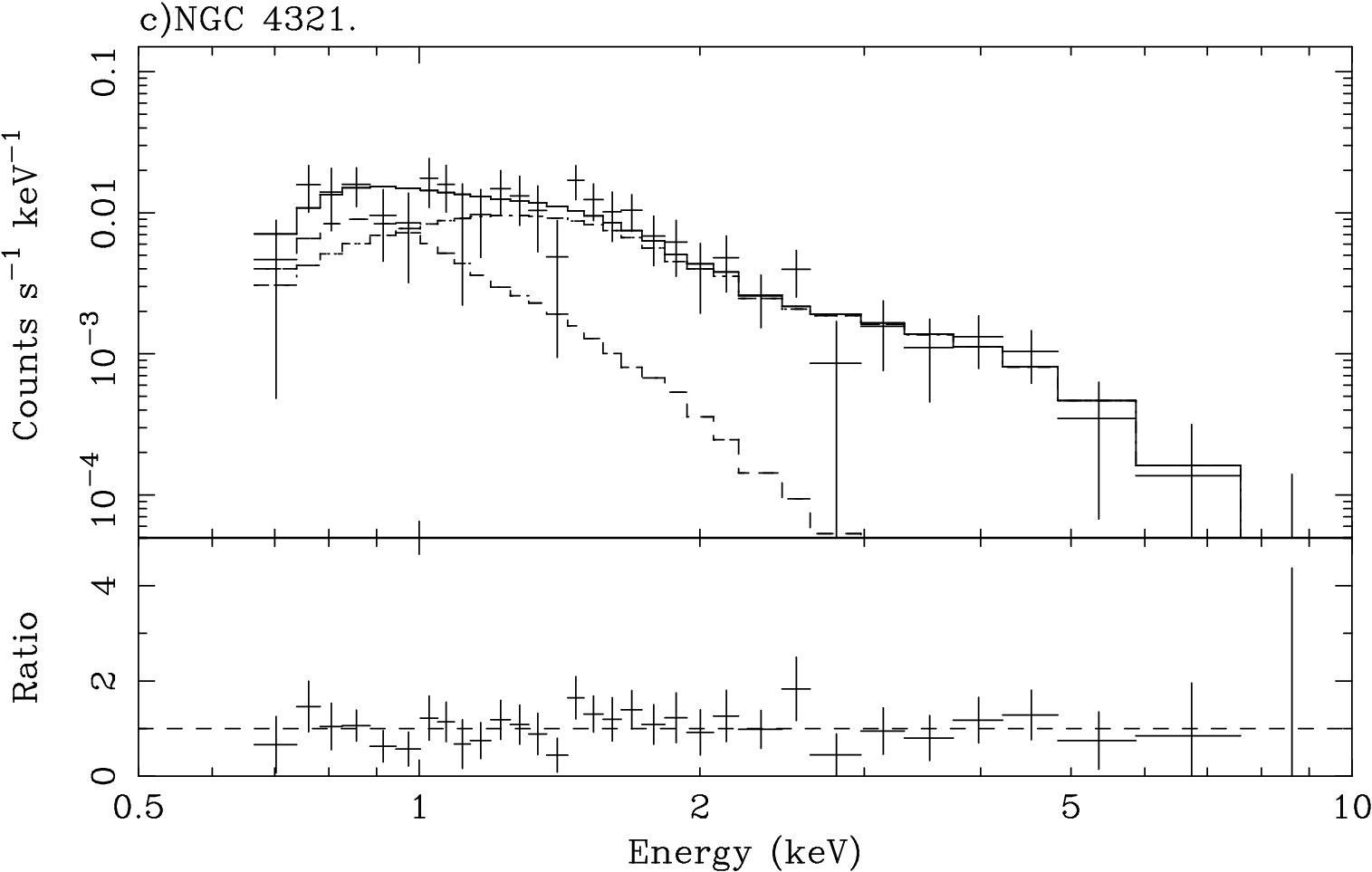}
\includegraphics[width=5.5 cm,angle=270]{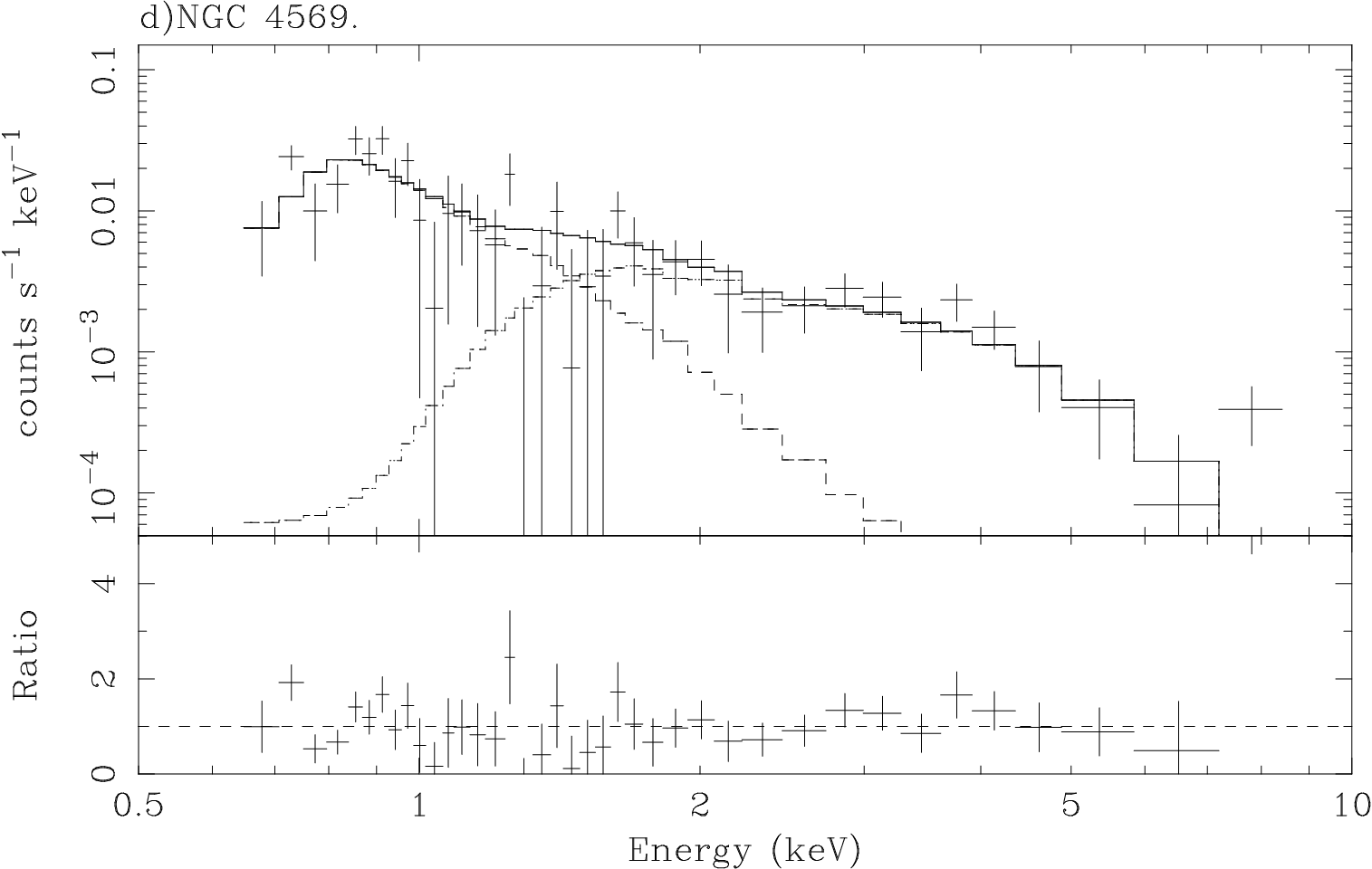}
}
\hbox{
\includegraphics[width=5.5 cm,angle=270]{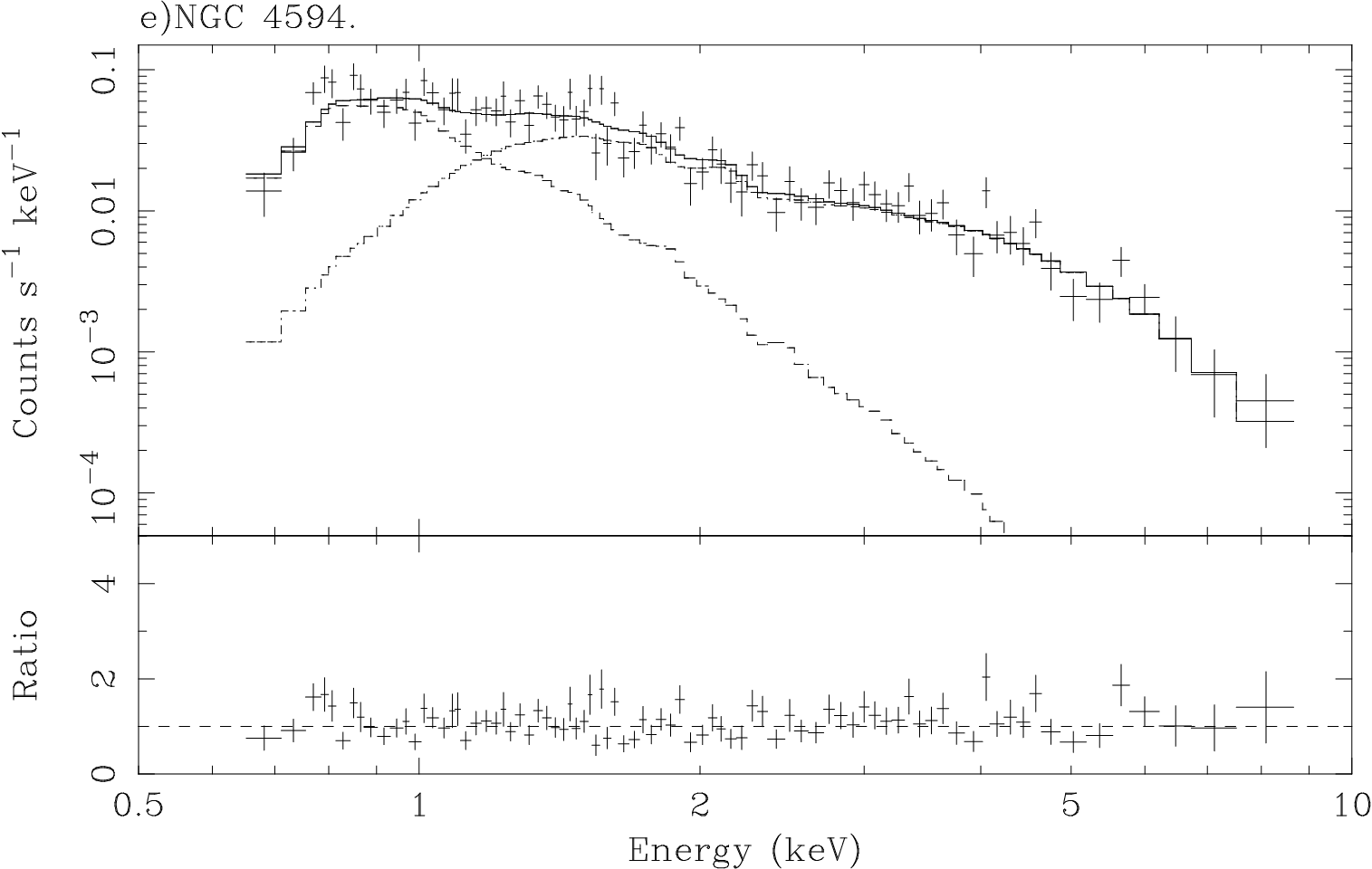}
\includegraphics[width=5.5 cm,angle=270]{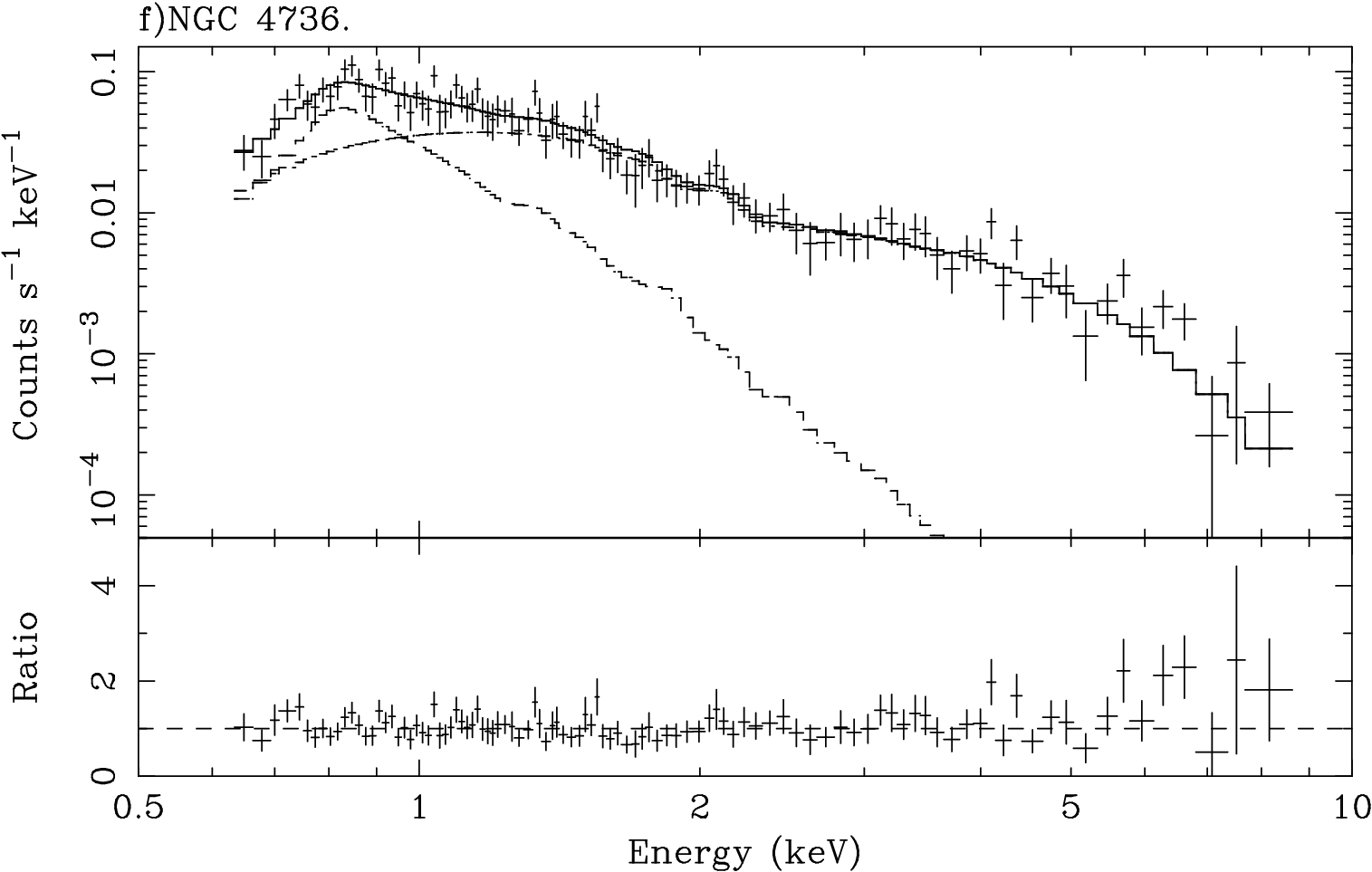}
}
\includegraphics[width=5.5 cm,angle=270]{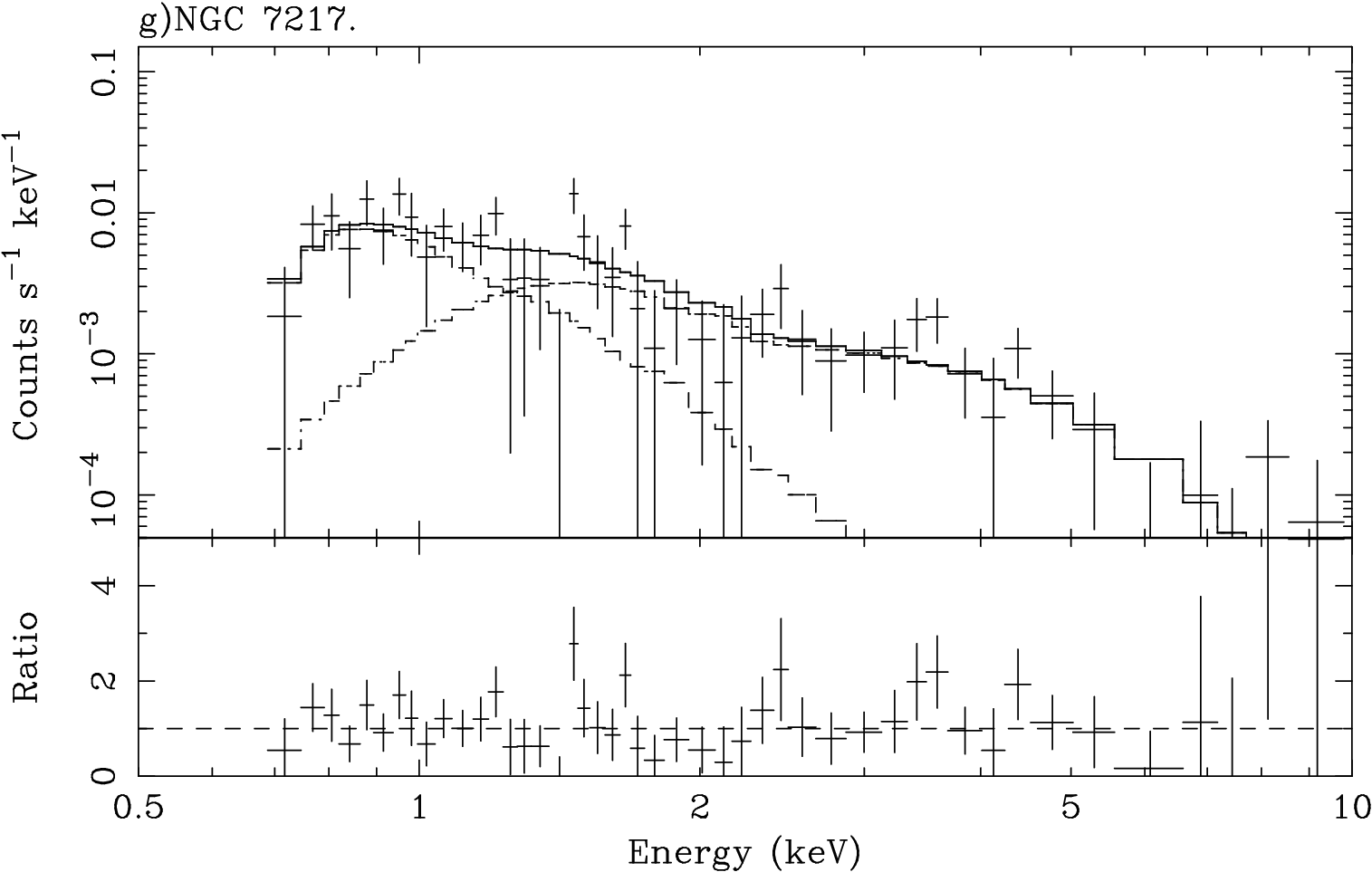}

\caption{The {\it ASCA\/} count rate spectra of the seven LINER 2 galaxies.
For clarity only the SIS 0 data are shown. In each case the lower panel
shows the ratio of the observed spectrum to the best fitting model.
}
\label{ascaspectra}
\end{figure*}

The spectral modelling was carried out using the XSPEC V10.0 software
package.  Data from all four \asca instruments were fit simultaneously
but with the relative model normalisation allowed to vary freely. The
quoted errors on the derived best-fitting model are based on a $\Delta
\chi^{2} = 2.71$ criterion, that is at a 90 per cent confidence level
for {\it one} interesting parameter.

The data for each source were initially fit with a model comprising a
power-law continuum of photon spectral index, $\Gamma$, with
low-energy absorption arising in a line-of-sight hydrogen column
density, \nh. In terms of the minimum $\chi^{2}$, this simple model
provided quite a reasonable description of observed spectra for four
of the sources (NGC 3627, NGC 3628, NGC 4321, NGC 7217) with the
measured values of $\Gamma$ ranging from 1.4 -- 2.3 and low values of
column density (\nh $<2\times10^{21}$), consistent with the Galactic
value in three of the four sources.

Ptak et al. (1999) have shown that the observed X-ray spectra of
low-luminosity galactic nuclei, whether driven by accretion or
starburst processes, can often be fitted by a ``standard spectral
model''. This model consists of soft X-ray emission attributed to a
Raymond-Smith plasma ($T_{RS} \la 10^{7}$ K), which contributes mostly
below $\sim 2$ keV, plus a harder emission component which can be
modelled either as a power-law ($\Gamma \approx 1.7$) or a thermal
bremsstrahlung ($kT \approx 5$ keV) continuum. The soft Raymond-Smith
component generally shows little evidence for absorption over and
above the line-of-sight Galactic column density whereas the hard
component often exhibits some additional intrinsic column density
(typically $N_{H,int} \sim 10^{22} \rm cm^{-2}$). In addition the
abundances inferred for the Raymond-Smith plasma are generally
significantly sub-solar, although this may be more indicative of a
complex multi-temperature plasma than a true under-representation of
elements such as iron and oxygen.

Given the relatively low count rates, and hence restricted
signal-to-noise, of the X-ray spectra for sources in our current
sample, we have applied some additional restrictions in our usage of
the above ``standard spectral model''. Thus the Raymond-Smith
component is assumed (i) to be subject only to absorption by the line
of sight Galactic column density, (ii) to have temperature $kT_{RS}$
in a restricted range of 0.6--0.9 keV (corresponding to the peak in
the temperature distribution measured by Ptak et al. 1999, see their
Fig.4) and (iii) to have a metal abundance fixed at 0.1 times the
solar value.  Table~\ref{bestfits} details the results from the
fitting of this standard spectral model. This standard model provides
an excellent match to the observed spectra and in all cases bar one,
it provides a very significant improvement in the minimum $\chi^{2}$
in compared to the simple absorbed power-law model (\ie
$\Delta\chi^{2}$ in Table~\ref{bestfits}). Table~\ref{bestfits} also
lists the observed 0.5-10 keV X-ray luminosity for each source
(corrected for Galactic absorption).  Fig.~\ref{ascaspectra} shows the
\asca count rate spectra and the corresponding best fitting model.

As a final step we have attempted to fit all the sources with an iron
K$_{\alpha}$ line.  This feature is seen in several LLAGN, though it
is a matter of debate whether it is indicative of X-ray reprocessing
in matter surrounding an active nucleus, as is the case in more
luminous AGN (see Terashima et al. 2000b, and references therein).
Given the restricted signal to noise we first add a narrow Gaussian
line component to the best-fitting model at a fixed energy.
Table~\ref{iron} summarises the limits on the line equivalent width of
lines placed at energies of 6.4 keV (corresponding to low ionization
states for iron) and 6.7 keV (helium-like iron).  When we then set the
line energy as a free parameter in the range 6.4-6.96 keV, there were
Fe K$_{\alpha}$ line detections in only two sources (NGC 4736 and NGC
7217).  However, both detections were marginal, most particularly in
the case of NGC 4736 (see Table~\ref{iron})

\section{Discussion}

Our objective in focussing this analysis on a relatively homogeneous,
albeit small, sample of early type spiral galaxies has been to
investigate whether there are any properties that the sample galaxies
have in common, which may be linked to their LINER 2
classification. In principle, if such a link could be established this
might provide some clue as to whether LINER 2 characteristics are
derived from the presence of a LLAGN or by some alternative shared
mechanism.

\subsection{Comparing \ro HRI and \asca observations}

As noted previously, the \ro HRI images reveal the presence of a
point-like X-ray source coincident with the optical nucleus in all
seven galaxies. The inferred soft X-ray luminosities of these
``nuclear'' X-ray sources (hereafter NXS) ranges from $1.0 - 45
\times 10^{39} \rm~erg~s^{-1}$ (Table 3). In most of the galaxies the
NXS contributes $\sim 50\%$ or more of the central galactic soft X-ray
emission.  Though the presence of a dominant NXS indicates a strong
soft X-ray emission peak coincident with the nucleus it is not, in
isolation, unambiguous evidence for a LLAGN. The $\sim 5''$ spatial
resolution of the HRI translates to a linear scale of 100--500 pc for
the sample galaxies, which covers the inner bulge component of each
galaxy and will encompass any nuclear and/or circumnuclear starburst
regions in addition to any putative active nucleus. Studies of normal
spiral and starburst galaxies have demonstrated that X-ray
luminosities of up to $10^{40} \rm~erg~s^{-1}$ can be generated
through the summed emission of the underlying galactic X-ray source
populations (high and low mass X-ray binaries, recent supernovae and
supernova remnants) and energetic diffuse processes (Fabbiano
1989). For most of our sample this can provide an alternative
explanation for the observed nuclear X-ray emission.  Clearly further
constraints are required based on either the temporal or spectral
properties of the NXS revealed by the
\ro observations.

The X-ray spectra measured by \asca for this set of galaxies show
remarkable similarities as evidenced by the fact that our
``standard'', thermal plus absorbed power-law, spectral model provides
a good fit in all seven cases.  Fig.~\ref{ascaspectra} illustrates the
relative contributions of the thermal and power-law components to the
observed count rate spectra (as derived from the spectral
fitting). Interestingly, the cross-over between the two spectral
components, for this set of sources, occurs within a very narrow
energy range, namely 0.9--1.5 keV. Since the spectral model implies
that the measured flux in the 0.5--1 keV band (corrected for Galactic
absorption) is largely attributable to thermal emission, whereas that
in the 2--5 keV band is attributable to the power-law component, it
follows that the ratio $F_X(0.5-1)/F_X(2-5)$ provides a measure of the
relative strength of the two components. This ratio is tabulated in
Table~\ref{ratios} for each galaxy. Excluding NGC 3628 and NGC 4594,
for which there is good existing evidence for the presence of a LLAGN
or, alternately in the case of NGC 3628, an ultra-luminous binary
dominant in the X-ray regime (see the notes in the Appendix), the five
remaining sources have a mean value for this ratio of 0.66 with a
standard deviation of only 0.10.  This commonality of spectral form
argues strongly that the 0.5--5 keV X-ray flux derives from the same
emission process (or a linked set of processes) in these five
galaxies.  This is a rather powerful argument {\it against} the LLAGN
hypothesis since there is no obvious reason why the soft thermal
emission should be so closely correlated with the level of the hard
power-law continuum emanating from a LLAGN.  In terms of the above
spectral ratio, the X-ray spectra of NGC 3628 and NGC 4594 are,
respectively, factors of 3.7 and 1.7 harder than the average for the
other five galaxies; it is plausible that this is due to the
contribution of an underlying LLAGN in these two galaxies.

It is notable that when $F_X(0.5-1)/F_X(2-5)$ is derived for the
classic starburst galaxy NGC 253, based upon the \asca data fits of
Ptak et al. (1999), an identical value of 0.66 is obtained.  Recent
analysis of the {\it XMM-Newton\/} PV observation of this galaxy
(Pietsch et al. 2001) has shown its X-ray emission to be dominated by
thermal (i.e. stellar) processes, with no evidence at all for a LLAGN
component.  It appears, then, that the $F_X(0.5-1)/F_X(2-5)$ of $\sim
0.66$ is readily explained by stellar processes alone.  We note that
this result cannot be corroborated by analysis of the other classic
starburst galaxy, M82, since on some occasions the X-ray emission of
this galaxy is dominated by a single, variable, luminous X-ray source 
(Matsumoto et al. 2001).

The \asca measurements sample a relatively large area of each galaxy
(a $4'$ radius region centred on the nucleus) with a harder spectral
response than that of the \ro HRI. As a result it is not
straightforward to relate the \asca spectral measurements to the
properties of the NXS identified in the \ro images. One approach is to
assume that the absorbed power-law emission originates solely in the
NXS. Using appropriate HRI response files in XSPEC, we have derived
the expected \ro HRI count rate for the power-law component based on
the spectral parameters and normalisations obtained from the \asca
measurements.  In all cases the predicted count rate is {\it less than
that actually observed} from the NXS.  Variability might, in
principle, give rise to some disparity between the \asca and \ro
measurements, but only in the case of NGC 3628 is there explicit
evidence for such behaviour (see the notes in the Appendix). This bias
towards the under-prediction of the NXS flux is perhaps explained if
at least some of the thermal emission modelled in \asca also
originates from the compact nuclear source.  

An alternative viewpoint would be to suggest that the hard power-law
emission does not originate solely in the nucleus.  This would
certainly appear to be consistent with the results of Terashima et
al.(2000a), who report that the hard X-ray emission associated with
several nearby LINER 2 nuclei is extended on scales of several kpc.
In this case, the soft X-ray NXS could be almost entirely dominated by
the thermal component.  This, again, supports the assertion that the
LINER 2 nuclei need not harbour LLAGN.  Unfortunately further
quantitative analysis of the two scenarios outlined above cannot be
justified given the mis-match of the \asca and \ro HRI bandpasses. For
example, \ro PSPC and HRI measurements indicate the presence of a very
soft thermal ($\sim 0.1$ keV) component in NGC 4736 which is not
evident in the \asca spectra (Roberts, Warwick \& Ohashi
1999). Clearly much more definitive analysis will be possible with
observations combining good spectral and excellent spatial resolution,
such as those from \chan.

\subsection{Luminosity ratio diagnostics}

\begin{table*}
\centering
\caption{X-ray flux and luminosity ratios of the LINER 2s.}
\begin{tabular}{lcccc}
Galaxy	& ${F_X(0.5-1)}/{F_X(2-5)}$	& log ${L_X}/{L_{B}}$ 
& log $L_X/L_{FIR}$ & log $L_X/L_{H\alpha}$  
\\
& & & & \\
NGC 3627	& 0.56	& -3.4	& -3.9	& 1.0 \\
NGC 3628	& 0.18	& -3.3	& -4.4	& 2.8 \\
NGC 4321	& 0.59	& -3.1	& -4.2	& 1.1 \\
NGC 4569	& 0.72	& -3.5	& -3.4	& -0.6 \\
NGC 4594	& 0.38	& -3.4	& -2.8	& 1.2 \\
NGC 4736	& 0.80	& -3.3	& -3.8	& 1.5 \\
NGC 7217	& 0.63	& -3.4	& -3.5	& 0.1 \\
\end{tabular}
\label{ratios}
\end{table*}

The ratio of the X-ray luminosity to the luminosity in other wavebands
can be a very useful diagnostic of the underlying emission processes
occurring within a galaxy. It has long been suspected that the hard
component present in the X-ray spectra of normal (i.e. non-active)
early-type galaxies is the signature of a ubiquitous population of
low-mass X-ray binaries (LMXBs), and that the X-ray luminosity of this
component is well correlated with the blue luminosity of the galaxy
(e.g. Canizares et al. 1987; Matsushita et al. 1994).  Iyomoto et
al. (1998) have recently used this relationship to demonstrate that
LLAGN are probably present in the elliptical galaxies NGC 3065 and NGC
4203, by showing that the ratio of the 2--10 keV X-ray luminosity
contained in a hard power-law component, to the galactic blue
luminosity, is two orders of magnitude greater for these
galaxies than in normal ellipticals (\cf Fig. 5 of Iyomoto et
al. 1998).  The same test can be applied to the present LINER 2 sample
using the blue luminosity of their galaxy bulges. The ratio $L_{X
(2-10, {\rm PL})}/L_{B,bulge}$ (hereafter $L_X/L_B$) is listed in
Table~\ref{ratios}. It shows remarkably little scatter throughout the
sample, especially when we consider that the inherent range in the
bulge blue luminosity is $\sim 1.6 - 60 \times 10^9 L_{\odot}$, \ie a
factor of $\sim 40$.  For the early-type galaxies in Iyomoto et al.,
typically log $L_X/L_B \approx -4.1$. Similarly Irwin \& Bregman
(1999) obtain a value log $L_X/L_B = -4.26$ for the bulge component of
M31.  The implication is that the LINER 2s studied here have an
observed X-ray luminosity which is a factor 4--10 higher than the
predicted LMXB contribution.  The excess X-ray emission may be
directly attributable to the presence of a LLAGN, and indeed the log
$L_X/L_B$ values in Table~\ref{ratios} are similar to those found for
{\it bona fide\/} LLAGN in spiral galaxies (Iyomoto et al. 1998).
However, given their later morphological type in comparison to the
galaxies studied by Iyomoto et al. (1998), an equally plausible
explanation is that our LINER 2 sample have nuclei hosting a more
energetic (and perhaps younger) X-ray binary population. The absence
of rapid (or, in fact, any) temporal variations in the hard X-ray
emission from our sample is consistent with the hard flux originating
in an unresolved X-ray binary population.

If the dominant contribution to the X-ray luminosity in our LINER 2
sample is attributable to processes associated with an underlying
young stellar population, then we might expect to see the X-ray
luminosity associated with the hot gas component scale with other
measures of star-formation activity. Terashima et al. (2000a) test for
this by comparing the intrinsic 0.5--4 keV X-ray luminosity of the hot
thermal component in spectra of several LINER 2 galaxies with the
far-infrared luminosity of the host galaxy, \ie by examining the ratio
$L_{X(0.5-4, {\rm RS})}/{L_{FIR}}$ (hereafter $L_X/L_{FIR}$).  Known
starburst galaxies have log $L_X/L_{FIR} \approx -4 \pm 1$ and, in
terms of this ratio, Terashima et al. (2000a) find that their LINER 2
galaxies have somewhat similar characteristics. The same comment also
applies to the present sample (see Table~\ref{ratios}, where we used
$L_{FIR}$ values published in HFS). The exception to this result is
NGC 4594 which has a notably higher X-ray to FIR ratio.  This anomaly
may be due to the presence of a hot ISM associated with the dominant
massive bulge component in NGC 4594.  We can conclude that, aside from
NGC 4594, the X-ray luminosity residing in the thermal component is at
a level consistent with a starburst origin.  Furthermore, the
constancy of spectral form noted in the previous section would imply
that the hard power-law component also has a starburst connection and
that this is the origin of the apparent excess of hard flux noted
above.

Terashima et al. (2000a) also discussed the \Ha~ionising photon budget
for low luminosity Seyfert galaxies and for LINERS.  Specifically they
investigate whether the underlying UV to X-ray continuum can produce 
the observed \Ha~flux, or whether additional ionising sources are
required.  Following a number of assumptions, they conclude that log
$L_X/L_{H\alpha} \ge 1.4$ is required to explain the \Ha~luminosity
solely in terms of photoionization from a LLAGN. The sample of LINER
2s considered by Terashima et al.(2000a) all have log $L_X/L_{H\alpha}
\le 0.6$ implying either the primary ionisation source is not a LLAGN
or that the line of sight to the LLAGN is heavy obscured in X-rays and
possibly Compton thick.  Derived values for the $L_X/L_{H\alpha}$
ratio for the present sample of LINER 2s are given in
Table~\ref{ratios} (where $L_X = L_{X(2-10,{\rm PL})}$ and $L_{H\alpha}$
is taken from HFS).  In the case of NGC 3628, the hard X-ray to
\Ha~luminosity ratio far exceeds the \Ha~ionisation requirement,
consistent with presence of a LLAGN. At the other extreme NGC 4569 and
NGC 7217 would appear to be under-luminous in X-rays in relation to
their observed \Ha~flux. The four remaining galaxies in our sample
have log $L_X/L_{H\alpha} = 1 - 1.5$, a somewhat indeterminate level
in terms of identifying the presence or absence of a LLAGN.

\section{Conclusion}

There is rather mixed evidence concerning the presence of a LLAGN in
the seven LINER 2 galaxies which comprise the present sample.  NGC
3628 is a highly variable X-ray source (Dahlem, Heckman \& Fabbiano
1995), has an X-ray spectrum that is dominated by a hard component and
has a high $L_X/L_{H\alpha}$ ratio, which may possibly originate in a
LLAGN, though it is also possible that its X-ray emission is dominated
by an ultra-luminous X-ray binary system. NGC 4594 also shows some
anomalous X-ray properties but displays the best observational
evidence for a LLAGN, which is provided by the recent discovery of
broad wings on its nuclear emission lines (Kormendy et al. 1996) plus
the incontrovertible case for a $10^{9}\Msun$ black hole in this
galaxy (Kormendy 1988).  At the other extreme, both NGC 4569 and NGC
7217 appear to be LINER 2 systems driven predominantly by stellar
processes, one manifestation of which is their low $L_X/L_{H\alpha}$
ratio.  For the three remaining sources, NGC 3627, NGC 4321 and NGC
4736 there is some evidence favouring the presence of a LLAGN (\eg the
radio source associated with the nucleus of NGC 3627; Filho, Barthel
\& Ho 2000) but this is far from overwhelming. However, the similarity
in spectral form (when the two reasonably well established
accretion-driven systems are excluded) with the predominantly stellar
driven systems (including NGC 253), noted in section 5.1, tips the
argument against the presence of a dominant LLAGN in these three LINER
2 nuclei.

The study of this small sample of LINER 2s has clearly advanced the
case for heterogeneity in the processes underlying the LINER 2
phenomenon. However, much more definitive analysis is required and
indeed will be possible with X-ray measurements which combine good
spatial resolution, reasonable spectral resolution and, ideally,
better temporal sampling.  The search for LLAGN in nearby galaxies of
all types is, of course, a high priority for both the \chan and
\xmm missions.

\vspace{1cm}

{\noindent \bf ACKNOWLEDGMENTS}

\vspace{2mm}

NJS and TPR gratefully acknowledge financial support from PPARC.  The
archival X-ray data used in this work were all obtained from the
Leicester Database and Archive Service (LEDAS) at the Department of
Physics \& Astronomy, University of Leicester.  The Digitised Sky
Survey was produced at the Space Telescope Science Institute, under US
government grant NAG W-2166 from the original National
Geographic--Palomar Sky Survey plates.  We thank the anonymous referee
for suggestions that have led to the improvement of this paper.


\appendix

\section{Notes on the sample galaxies}

\subsection{NGC 3627}

NGC 3627 (M 66) is an interacting galaxy in the Leo triplet.  HFS
classify it as T2/S2 type, indicating some ambiguity in
its optical emission-line ratios leading to doubt as to whether the
nucleus is a LINER/\hii composite or Seyfert 2-like (with the former
being the most probable option).  Filho, Barthel \& Ho (2000) report
the presence of a variable, flat spectrum radio source in the nucleus
of NGC 3627 consistent the presence of a LLAGN.

NGC 3627 was detected with a count rate of $\sim 0.1 \ctsec$ in the 
\ro All-Sky Survey, but the only previously reported observation is from 
a \ro PSPC image of NGC 3628, where NGC 3627 is 36$'$ off-axis (Dahlem
et al 1996).  The strong off-axis degradation of the instrument point
spread function prevented any useful spatial analysis, but spectral
analysis was possible. The data were reasonably well fitted by a
combined thermal plasma and power-law continuum model ($kT = 0.34 \pm
0.69$ keV, $\Gamma = 2.0 \pm 0.59$ respectively) with no evidence for
absorption above the Galactic line-of-sight value.  The inferred 0.1
-- 2.0 keV luminosity was $5 \times 10^{39} \ergsec$ for an assumed
distance of 6.6 Mpc.

\subsection{NGC 3628}

NGC 3628 is an edge-on galaxy, also in the Leo Triplet, that shows
clear morphological distortions (e.g. in \hi; Haynes et al. 1979) due
to its interaction with the other Leo Triplet galaxies.  Although the
nucleus is obscured by a prominent dust lane, there is evidence for a
nuclear starburst, though it may be intrinsically less active than in
the prototypical starburst galaxies NGC 253 and M 82 (see Dahlem et
al. 1996 and references therein).

The presence of the nuclear starburst has made NGC 3628 a popular
target for X-ray observations.  \ein observations of the galaxy
detected two point X-ray sources, one of which was coincident with the
nucleus (Bregman \& Glassgold 1982), together with an X-ray bright
outflow aligned along the minor axis of the galaxy (Fabbiano, Heckman
\& Keel 1990). The presence of the latter was confirmed by \ro PSPC
observations (Dahlem et al. 1996).  The nuclear source has shown a
remarkable level of variability, with its flux changing by factors of
up to $\sim 20$ between observations. This establishes the domination
of a single source which may be either a LLAGN or luminous X-ray
binary (Dahlem, Heckman \& Fabbiano 1995).  If it is a luminous
binary, then its luminosity is such that it is probably an example of
the so-called ultra-luminous X-ray source phenomena (ULX; Makishima et
al. 2000) situated close to the dynamical centre of the galaxy, as is
seen in many nearby galaxies (Colbert \& Mushotzky 1999).  Spectral
studies of NGC 3628 using \asca have shown the nuclear source to have
a very flat spectral shape, with simple powerlaw fits giving a
spectral index $\Gamma \approx 1.2$ (Yaqoob et al.  1995). More
complicated models (including one or more thermal plasma components
and a heavily-absorbed power-law continuum) have also been
successfully fitted to the \asca and \ro PSPC data (Dahlem, Weaver \&
Heckman 1998; Ptak et al. 1999).

\subsection{NGC 4321}

A remarkable feature of the well-known grand design spiral NGC 4321 (M
100) is the number of recent supernovae in the galaxy, numbering 4 in
the last 100 years.  The \ein observatory targeted the most recent,
SN1979C, two years after its appearance, but it did not detected the
SN in X-rays (Palumbo et al. 1981).  Instead, \ein saw a bright
nuclear source and a second source in the northern spiral arm.
Immler, Pietsch \& Aschenbach (1998) report the results of a \ro HRI
observation of NGC 4321 which detects several point sources within the
\d25 ellipse of the galaxy, including the bright nuclear source (which
has a luminosity of \lx $\approx 6 \times 10^{39} \ergsec$) and SN
1979C (at a luminosity of \lx $= 1 \times 10^{39} \ergsec$), 16 years
after its initial outburst.  NGC 4321 has not been studied
spectroscopically in X-rays to date.

\subsection{NGC 4569}

NGC 4569 (M 90) is in the Virgo cluster.  It has a very bright
nucleus, probably the result of a recent star formation episode
(Stauffer, Kenney \& Young 1996).  Optical and {\it IUE\/} studies
(Keel 1996) have demonstrated that the nucleus is the host to a young star
cluster, dominated optically by A-type supergiants.  \hst UV imaging
showed the nucleus to be a bright, point-like, UV source (Maoz et
al. 1995), although more recent observations indicate that the UV
source is clearly extended (Barth et al. 1998).  UV spectroscopy
reveals that the dominant continuum source is a cluster of massive
stars and that there is no evidence in the UV for a LLAGN
(Maoz et al. 1998). Narrow-band ([O III]$\lambda 5007$ and H$\alpha+$
[N II]) \hst optical imaging on the other hand, shows the nucleus to
be bright and unresolved on a sub-arcsecond scale (Pogge et al. 2000).

A \ro PSPC observation reveals extended X-ray emission in the bulge and
disk of NGC 4569, with the emission peaking at the nucleus (Junkes
\& Hensler 1996).  The HRI observation, however, shows a luminous, 
unresolved point source coincident with the nucleus (Colbert
\& Mushotzky 1999; Roberts \& Warwick 2000).  \asca observations of
NGC 4569 confirm that the X-ray emission is extended on scales of
several kpc in both the soft (0.5 - 2 keV) and hard (2 - 7 keV) \asca
bands (Terashima et al. 2000a).  The \asca spectra are best described
by a Raymond-Smith plus power-law (or thermal bremsstrahlung) model,
with parameters $kT \approx 0.67$ and $\Gamma \approx 2.2$
respectively; an additional column density of $1.7 \times 10^{21}
\atpcm$ applies to the second component.  Terashima et al. (2000a)
postulate that there is no evidence for the presence of an AGN in NGC
4569; the soft component could originate in star formation, whereas
the hard component could easily be the sum of a discrete (non-AGN)
source contribution.  In combination with the other measurements
summarised above, Terashima et al. (2000a) conclude that the LINER
spectrum of NGC 4569 is the probably the result of the presence of hot
stars in its nucleus.

\subsection{NGC 4594}

The famous galaxy NGC 4594 (M 104, the Sombrero) was one of the
earliest galaxies to show evidence for the possible presence of a
super-massive (up to $10^{9} M_{\odot}$) black hole in its nucleus
(Kormendy 1988).  It was classified by HFS as having a LINER 2
nucleus, and there is an abundance of evidence to suggest that it is
powered by a LLAGN.  Most convincingly, the high-resolution \hst
spectrum of NGC 4594 shows broad wings on the nuclear emission lines
(Kormendy et al. 1996).Examination of the ionization budget for the
optical lines highlights a deficit of near-UV ionizing photons which
points to a non-stellar component in the nucleus (Maoz et al. 1998),
as does the FIR-radio ratio when compared to Wolf-Rayet galaxies (Ji
et al. 2000).  Intriguingly, the SED of the nucleus does not show
evidence for a ``big blue bump'' as seen in more luminous AGN,
although this may be an implicit characteristic of the LLAGN class (Ho
1999).

The X-ray emission of NGC 4594 has been very well studied.  \ein
showed an extended X-ray source to be centred on the nucleus of the
galaxy (Fabbiano, Kim \& Trinchieri 1992).  Fabbiano \& Juda (1997)
investigated a \ro HRI observation, finding that the emission broke
down into three components: a bright, point-like nuclear source with
\lx $\sim 3.5 \times 10^{40} \ergsec$ (0.1 - 2.4 keV), clumpy disk
emission and unresolved bulge component.  They concluded that if the
nuclear emission is due to accretion onto a super-massive black hole,
then it must be severely sub-Eddington.  \asca spectroscopy was
investigated by Nicholson et al. (1998); the best fit to the combined
\asca and \ro PSPC data is a power-law continuum with $\Gamma \approx
1.6$ and \nh~slightly above the Galactic value. However, if a putative
soft X-ray cut-off is masked by the extended thermal emission component 
then the absorption on the power-law component could rise to $3 \times
10^{21} \atpcm$.  Nicholson et al. also look at \hst UV spectra, and
find the properties of NGC 4594 are consistent with the presence of a
LLAGN.  Ptak et al. (1999) also fit a 2-component model to the \asca
and PSPC data, with similar results except that they find more
intrinsic absorption on the powerlaw ($\sim 10^{22}\atpcm$).

\subsection{NGC 4736}

NGC 4736 (M 94) hosts the closest example of a LINER 2 nucleus, and so
is amongst the best studied.  There is conflicting evidence as to
whether the LINER 2 in NGC 4736 is powered by a LLAGN. Larkin et al. 
(1998) conducted a near-infrared spectroscopic survey of
LINER galaxies and found NGC 4736 to have the largest [Fe
II]/Pa$\beta$ ratio in their sample, indicating that its excitation
was dominated by stellar processes. Both the H$\alpha$ luminosity
(Taniguchi et al. 1996) and FIR emission (Smith et al. 1991; Smith et
al. 1994) may also be explained solely by stellar emission.  On the
other hand, Turner \& Ho (1994) detect a strong non-thermal radio
continuum source at the position of the nucleus.  Maoz et al. (1995)
image the nucleus in UV using the \hst, and detect {\it two\/} bright
point sources in the nuclear region and possible bow shocks, leading
to the conjecture that two objects are in the final stages of merging
in the nucleus.  One or both of these sources may be very compact star
clusters, but they may also be super-massive black holes, one of which
comes from a merged, nucleated satellite galaxy (as postulated in
Taniguchi \& Wada 1996).

Cui et al. (1997) studied a \ro PSPC observation of NGC 4736 which
revealed a bright, nuclear X-ray source surrounded by an extended
region of soft ($kT \approx 0.3$ keV) thermal gas.  The PSPC data was
re-analysed, along with two \ro HRI observations and an \asca
observation of the galaxy, by Roberts, Warwick \& Ohashi (1999).  They
found the X-ray spectrum to be dominated by a Seyfert-like power-law
continuum with slope $\Gamma \approx 1.7$, with little intrinsic
absorption. Below 2 keV two thermal emission components were required
with $kT$ in the range 0.1 - 0.6 keV. The possible detection of an Fe
K$\alpha$ feature at 6.8 keV was also reported. The compact nuclear
source was found to have a luminosity of $6 \times 10^{39} \ergsec$
(0.5 - 10 keV) and thus may be a near-quiescent AGN. However, the
alternative hypothesis that the hard component originates entirely in
X-ray binaries could not be ruled out.

\subsection{NGC 7217}

NGC 7217 is one of the most spheroid-dominated spiral galaxies known,
containing a series of ring structures (Buta et al. 1995).  Most
notable of these is a central dust ring, $\sim 17''$ in diameter,
around a dust-free nucleus.  No UV emission was detected from the
nucleus by Barth et al. (1998).  The NIR spectroscopic survey of
Larkin et al. (1998) found the nucleus to have a high [Fe
II]/Pa$\beta$ ratio, and therefore to be dominated by stellar
processes, a conclusion which is backed up by the FIR-radio relation
of Ji et al. (2000).This galaxy has not been previously studied in X-rays.


\begin{thebibliography}{}

\bibitem[]{} 
Barth A.J., Ho L.C., Filippenko A.V., Sargent W.L.W., 1998, ApJ, 496, 133
\bibitem[]{}
Bregman J.N., Glassgold A.E., 1982, ApJ, 263, 564
\bibitem[]{}
Buta R., van Driel W., Braine J., Combes F., Wakamatsu K., Sofue Y.,
Tomita A., 1995, ApJ, 450, 593
\bibitem[]{}
Canizares C.R., Fabbiano G., Trinchieri G., 1987, ApJ, 312, 503
\bibitem[]{}
Colbert E.J.M., Mushotzky R.F., 1999, ApJ, 519, 89
\bibitem{}
Cui W., Feldkhun D., Braun R., 1997, ApJ, 477, 693
\bibitem[]{} 
Dahlem M., Heckman T.M., Fabbiano G., 1995, ApJ, 442, L49
\bibitem[]{} 
Dahlem M., Heckman T.M., Fabbiano G., Lehnert M.D., Gilmore D., 1996,
ApJ, 461, 724
\bibitem[]{} 
Dahlem M., Weaver K.A., Heckman T.M., 1998, ApJS, 118, 401
\bibitem[]{} 
Fabbiano G., 1989, ARA\&A, 27, 87
\bibitem[]{} 
Fabbiano G., Heckman T., Keel W.C., 1990, ApJ, 355, 442
\bibitem[]{} 
Fabbiano G., Juda J.Z., ApJ, 1997, 476, 666
\bibitem[]{} 
Fabbiano G., Kim D.-W., Trinchieri G., 1992, ApJS, 80, 531
\bibitem[]{} 
Filho M.E., Barthel P.D., Ho L.C., 2000, ApJS, 129, 93
\bibitem[]{} 
Haynes M.P., Giovanelli R., Roberts M.S., 1979, ApJ, 229, 83
\bibitem[]{} 
Heckman T.M., 1980, A\&A, 87, 152
\bibitem[]{} 
Ho L.C., 1999, ApJ, 516, 672
\bibitem[]{} 
Ho L.C., Filippenko A.V., Sargent W.L.W., 1997, ApJS, 112, 315
\bibitem[]{} 
Immler S., Pietsch W., Aschenbach B., 1998, A\&A, 331, 601
\bibitem[]{} 
Irwin J.A., Bregman J.N., 1999, ApJ, 527, 125
\bibitem[]{} 
Iyomoto N., Makishima K., Matsushita K., Fukazawa Y., Tashiro M.,
Ohashi T., 1998, ApJ, 503, 168
\bibitem[]{} 
Ji L., Chen Y., Huang J.H., Gu Q.S., Lei S.J., 2000, A\&A, 355, 922
\bibitem[]{} 
Junkes N., Hensler G., 1996, in Zimmerman H.U., Tr{\"u}mper J., Yorke
H., eds.  {\it Proc. Rontgenstrahlung from the Universe}, MPE report
263, 459
\bibitem[]{} 
Keel W.C., 1996, PASP, 108, 917
\bibitem[]{} 
Kormendy J., 1988, ApJ, 335, 40
\bibitem[]{} 
Kormendy J., Bender R., Ajhar E.A., Dressler A., Faber S.M., Gebhardt
K., Grillmair C., Lauer T.R., Richstone D., Tremaine S., 1996, ApJL,
473, 91
\bibitem[]{} 
Larkin J.E., Armus L., Knop R.A., Soifer B.T., Matthews K., 1998,
ApJS, 114, 59
\bibitem[]{} 
Lawrence A., 1999, in Schmitt H.R., Kinney A.L., Ho L.C., eds.,
Adv. in Sp. Res., Vol. 23, No. 5/6, 'The AGN/normal galaxy
connection', pg. 1167, Oxford:Elsevier
\bibitem[]{} 
Makishima K., Kubota A., Mizuno T., Ohnishi T., Tashiro M., Aruga Y.,
Asai K., Dotani T., Mitsuda K., Ueda Y., Uno S., Yamaoka K., Ebisawa
K., Kohmura Y., Okada K., 2000, ApJ, 535, 632
\bibitem[]{} 
Maoz D., Filippenko A.V., Ho L.C., Rix H.-W., Bahcall J.N., Schneider
D.P., Macchetto F.D., 1995, ApJ, 440, 91
\bibitem[]{} 
Maoz D., Koratkar A., Shields J.C., Filippenko A.V., Sternburg A.,
1998, AJ, 116, 55
\bibitem[]{} 
Matsumoto H., Tsuru T.G., Koyama K., Awaki H., Canizares C.R., Kawai
N., Matsushita S., Prestwich A., Ward M., Zezas A.L., Kawabe R., 2001, 
ApJ, in press
\bibitem[]{} 
Matsushita K., Makishima K., Awaki H., Canizares C.R., Fabian A.C.,
Fukazawa Y., Loewenstein M., Matsumoto H., Mihara T., Mushotzky R.F.,
Ohashi T., Ricker G.R., Serlemitsos P.J., Tsuru T., Tsusaka Y.,
Yamasaki T., 1994, ApJL, 436, 41
\bibitem[]{} 
Nicholson K.L., Reichert G.A., Mason K.O., Puchnarewicz E.M., Ho L.C., 
Shields J.C., Filippenko A.V., 1998, MNRAS, 300, 893
\bibitem[]{} 
Palumbo G.G.C., Maccacaro T., Zamorani G., Panagia N., Vettolani G.,
1981, ApJ, 247, 484
\bibitem[]{} 
Pietsch W., Roberts T.P., Sako M., Freyberg M.J., Read A.M., Borozdin
K.N., Branduardi-Raymont G., Cappi M., Ehle M., Ferrando P., Kahn
S.M., Ponman T.J., Ptak A., Shirey R.E., Ward M., 2001, A\&A, in press
\bibitem[]{} 
Pogge R.W., Maoz D., Ho L.C., Eracleus M., 2000, ApJ, 532, 323
\bibitem[]{} 
Ptak A., Serlemitsos P., Yaqoob T., Mushotzky R., 1999, ApJS, 120, 179
\bibitem[]{} 
Ptak A., Serlemitsos P., Yaqoob T., Mushotzky R., Tsuru T., 1997, AJ,
113, 1286 
\bibitem[]{} 
Roberts T.P., Warwick R.S., 2000, MNRAS, 315, 98
\bibitem[]{} 
Roberts T.P., Warwick R.S., Ohashi T., 1999, MNRAS, 304, 52
\bibitem{}
Shields J.C., 1992, ApJL, 399, 27
\bibitem{}
Smith B.J., Lester D.F., Harvey P.M., Pogge R.W., 1991, ApJ, 373, 66
\bibitem{}
Smith B.J., Harvey P.M., Colom{\'e} C., Zhang C.Y., DiFrancesco
J., Pogge R.W., 1994, ApJ, 425, 91
\bibitem[]{} 
Stauffer J.R., Kenney J.D., Young J.S., 1986, AJ, 91, 1286
\bibitem{}
Taniguchi Y., Wada K., 1996, ApJ, 469, 581
\bibitem{}
Taniguchi Y., Ohyama Y., Yameda T., Mouti H., Yoshida M., 1996, ApJ,
467, 215
\bibitem[]{} 
Terashima Y., Ho L.C., Ptak A.F., Mushotzky R.F., Serlemitsos P.J.,
Yaqoob T., Kuneida H., 2000a, ApJ, 533, 729
\bibitem[]{} 
Terashima Y., Ho L.C., Ptak A.F., Yaqoob T., Kuneida H., Kazutami M.,
Serlemitsos P.J., 2000b, ApJL, 535, 79
\bibitem[]{} 
Terlevich R., Melnick J., 1985, MNRAS, 213, 841
\bibitem{}
Turner J.L., Ho P.T.P., 1994, ApJ, 421, 122
\bibitem[]{} 
Yaqoob T., Serlemitsos P.J., Ptak A., Mushotzky R., Kuneida H.,
Terashima Y., 1995, ApJ, 455, 508

\end{thebibliography}
\end{document}